\documentclass[runningheads]{llncs}
\usepackage{graphicx}
\usepackage[caption=false]{subfig}
\newcommand{\code}[1]{\texttt{#1}}

\begin{document}
\title{Ensemble of 3D CNN regressors with data fusion for fluid intelligence prediction }

\author{
Marina Pominova\inst{1} \and Anna Kuzina\inst{1} \and Ekaterina Kondrateva\inst{1} \and Svetlana Sushchinskaya\inst{1} \and
Maxim Sharaev\inst{1} \and Evgeny Burnaev\inst{1} \and Vyacheslav Yarkin\inst{1}}
\authorrunning{M. Pominova, A. Kuzina, E. Kondrateva et al.}
%
\institute{Skolkovo Institute of Science and Technology, Moscow, Russia
\email{ekaterina.kondrateva@skoltech.ru}\\}

\maketitle 
%

%
\begin{abstract}
In this work, we aim at predicting children's fluid intelligence scores based on structural T1-weighted MR images from the largest long-term study of brain development and child health. The target variable was regressed on a data collection site, sociodemographic variables and brain volume, thus being independent to the potentially informative factors, which are not directly related to the brain functioning. We investigate both feature extraction and deep learning approaches as well as different deep CNN architectures and their ensembles. We propose an advanced architecture of VoxCNNs ensemble, which yield MSE (92.838) on blind test. 

 \and VoxCNN ensemble 
\end{abstract}
\section{Introduction}



Understanding cognitive development in children may potentially improve their health outcomes through adolescence. Thus, determining neural mechanism underlying general intelligence is a critical task. One of two discrete factors of general intelligence is fluid intelligence.

Fluid intelligence is the capacity to think logically and solve problems in novel situations, independent of acquired knowledge. It involves the ability to identify patterns and relationships that underpin novel problems and to extrapolate these findings using logic \cite{carroll_1993}.

There are research devoted on fluid intelligence prediction based on different brain imaging techniques and extracted features \cite{zhu2018prediction},\cite{paul2016dissociable}. However, the authors could not highlight robust biomarkers and methods to predict fluid intelligence scores . 

Deep learning approaches and convolutional neural networks, in particular, have shown high potential on imagery classification, recognition and processing and thus could be considered useful for fluid intelligence scores prediction based on MRI data (3D brain images).

The advantage of deep learning methods is the ability to automatically derive complex and informative features from the raw data during the training process. That allows training a neural network directly on high-dimensional 3D brain imaging data skipping the feature extraction step.

By design, neural architectures for deep learning are built in a modular way, with basic building blocks, such as composite convolutional layers, typically reused across many models and applications. This enables the standardization of deep learning architectures, with much research devoted to the exploration of pre-built layers and pre-trained activations (for transfer learning, image retrieval, etc.). However, the choice of appropriate architecture targeting specific clinical applications such as cognitive potential prediction or pathology classification remains open problem and requires further investigation.

In the present study we carry out an extensive experimental evaluation of deep voxelwise neural network architectures for fluid intelligence scores prediction based on MRI data with multimodal input structure.

The article has the following structure. In Section \ref{overview} we overview deep network architectures used for MRI data processing. In Section \ref{materials} we present the training dataset and our deep network architecture.  We describe obtained results in Section \ref{results}, provide discussions in Section \ref{discussion} and draw conclusions in Section \ref{conclusion}.

\section{Related work}
\label{overview}

There is a number of successful applications of convolutional neural networks (CNN) with different architectures for segmentation of  MRI data. Many of these solutions are based on adapting existing approaches to analyzing 2D images for processing of three-dimensional data. 

For example, for segmentation of the brain, an architecture similar to ResNet \cite{he2016deep} was proposed, which expands the possibilities of deep residual learning for processing volumetric MRI data using 3D filters in convolutional layers. The model, called VoxResNet \cite{chen2018voxresnet}, consists of volumetric residual blocks (VoxRes blocks), containing convolutional layers as well as several deconvolutional layers. The authors demonstrated the potential of ResNet-like volumetric architectures, achieving better results than many modern methods of MRI image segmentation \cite{milletari2016v}. Convolutional neural networks also showed good classification results in problems associated with neuropsychiatric diseases such as Alzheimer's disease. 

Recently proposed classification model with a VGG-like architecture called VoxCNN was used for neuro-degenerative decease classification \cite{hosseini2016alzheimer}. These results were more accurate or comparable to earlier approaches that use previously extracted morphometrical lower dimensional brain characteristics \cite{Pipeline2018,Epilepsy2018,DepressionAWE2018}.

Thus, this indicates that convolutional networks can be applied directly to the raw neuroimaging data without loss of model performance and over-fitting, which allows skipping the pre-processing step.

However, to the depth of our knowledge, there has not been much work on the use of convolutional networks for predicting fluid intelligence based on MRI imaging.

\section {Materials and Methods}
\label{materials}

\subsection{Data set}

The training data set is provided by ABCD Neurocognitive Prediction Challenge (ABCD-NP-Challenge 2019\footnote[1]{https://sibis.sri.com/abcd-np-challenge/}). The data contained of T1-weighed MRI images for four thousand individuals (of age 9-10 years) and corresponding sociodemographic variables \cite{hagler2018image}. The participants' fluid intelligence scores  (4154 subjects, 3739 for training and 415 for validation) are also provided.

\subsection{Target processing} 

The fluid intelligence scores were pre-residualized on a data collection site, sociodemographic variables and brain volume. For that a linear regression model was fitted with fluid intelligence as the dependent variable and brain volume, data collection site, age at baseline, sex at birth, race/ethnicity, highest parental education, parental income, and parental marital status as independent variables \cite{hagler2018image}.

The obtained residuals are used as targets to be predicted by a regression model.

\subsection{MRI data processing}

Imagery dataset consists of skull stripped images affinely aligned to the SRI 24 atlas \cite{rohlfing2010sri24}, segmented into regions of interest according to the atlas, and the corresponding volume scores of each ROI \cite{pfefferbaum2017altered}. T1-weighted MRI was transformed according to the Minimal Processing Pipeline by ABCD \cite{hagler2018image}.

The cross-sectional component of the National Consortium on Alcohol and NeuroDevelopment in Adolescence (NCANDA) pipeline \cite{brown2015national} was applied to T1 images. The steps included noise removal and field inhomogeneity correction confined to the brain mask, defined by non-rigidly aligning SRI24 atlas to the T1w MRI via Advanced normalization tools (ANTS) \cite{avants2009advanced}.

The brain mask was refined by majority voting across maps extracted by FSL BET \cite{smith2002fast}, AFNI 3dSkullStrip \cite{cox1996afni}, FreeSurfer mrigcut \cite{sadananthan2010skull}, and the Robust Brain Extraction (ROBEX) methods \cite{iglesias2011robust}, which were applied on combinations of bias and non-bias corrected T1w images. Using the refined masked, image inhomogeneity correction was repeated and the skull-stripped T1w image was segmented into brain tissue (gray matter, white matter, and cerebrospinal fluid) via Atropos \cite{avants2011open}. Gray matter tissue was further parcelled according to the SRI24 atlas, which was non-rigidly registered to the T1w image via ANTS.

\subsection{Specifications of the investigated models}

We use an ensemble of deep neural networks with VoxCNN architecture \cite{korolev2017residual,pominova2018voxelwise} to solve the regression problem. The proposed architecture has already demonstrated some successful applications to brain image analysis tasks. To provide better convergence and stronger regularization of results we enhanced this architecture.

VoxCNN networks are similar to VGG \cite{simonyan2014very} architecture, which is a popular architecture for 2D-images classification. VoxCNN applies 3D convolutions to deal with three-dimensional MRI brain scans. 

Proposed network consists of four blocks with two convolutional layers each having 3D convolutions followed by batch-normalization and ReLU activation function \cite{eckle2019comparison}. Number of filters in convolutional layers starts from 16 in the first block and doubles with each next block. Filters of the very first layer are applied with the stride \code{x2} to reduce the dimension of the original image. Our experiments have shown that this step does not reduce the network performance but helps to speed up the convergence and meet the limitations of GPU memory. The blocks are separated by max-pooling layers. We also apply 3D-dropout after each pooling layer to promote independence between feature maps and reduce over-fitting \cite{tompson2015efficient}.

Next, feature maps extracted by the convolutional layers are fed into the fully connected layer with \code{1024} hidden units, batch-normalization, ReLU activation, and dropout regularization, and then to the final layer with a single unit without non-linearity.

It was previously shown that auxiliary tower backpropagates the classification loss earlier in the network, serving as an additional regularization mechanism \cite{szegedy2015going,szegedy2016rethinking}. 

Therefore, the auxiliary output was added to the network to provide better training of the deeper layers. For this purpose, feature maps from intermediate layers are fed to the separate fully connected layer to produce another target prediction, which is then added to the main network output with adjusted weight. In this case, the output of the third block of convolutional layer was used to compute auxiliary prediction and average it with the main output with weights \code{0.4} and \code{0.6} respectively.

 We estimate quality of the models by Mean Squared Error (MSE) between the predicted scores and the pre-residualized fluid intelligence scores. The models were selected by optimizing the \code{MSE-loss} with the Adam optimizer. The learning rate was set to \code{3e-5}, batch size is \code{10} and each network was trained until the loss on validation set starts to increase.

To train the model we use multi-modal input data: brain scan data (T1-weighted imagery after preprocessing) and gray matter segmented brain masks. For each subject, two three-dimensional images were stacked as channels of a single image. 
We fed the resulted 3D image with two channels into the \code{VoxCNN} network as an input.

We use cross-validation to increase the model performance: we divide the training sample into two separate parts and two neural networks are trained with the same architecture on each part independently. Then for the validation subjects, an ensemble of these two models, defined as a weighted average of their predictions, is applied. Weights for averaging are determined based on the validation performance of each model (test predictions of the network that turned out to demonstrate lower \code{MSE} score on validation were set to larger weights). The number of layers, \code{Stride} and \code{ReLU} blocks position were adjusted correspondingly.

The train set consists of \texttt{n = 3739} samples, the validation set -- \texttt{n = 415} samples, and the test set -- \texttt{n = 4515} samples.

The models were implemented in \code{PyTorch} and trained on a single GPU \cite{canziani2016analysis}.

\begin{figure}[t!]
    \centering
    \subfloat[Straight-forward\label{fig:label:a}]{
        
        \includegraphics[height=8cm, width=3cm]{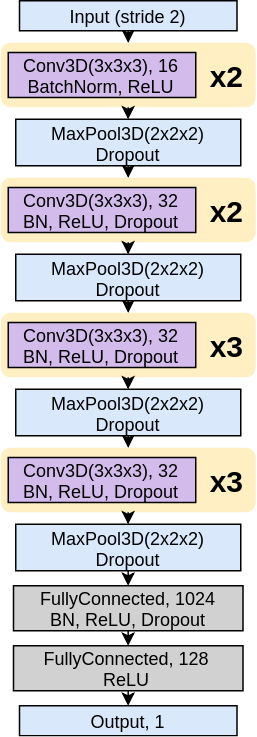}
        }
    \centering
    \subfloat[Architecture with auxiliary output\label{fig:label:b}]{
        \centering
        \includegraphics[height=8cm,width=6cm]{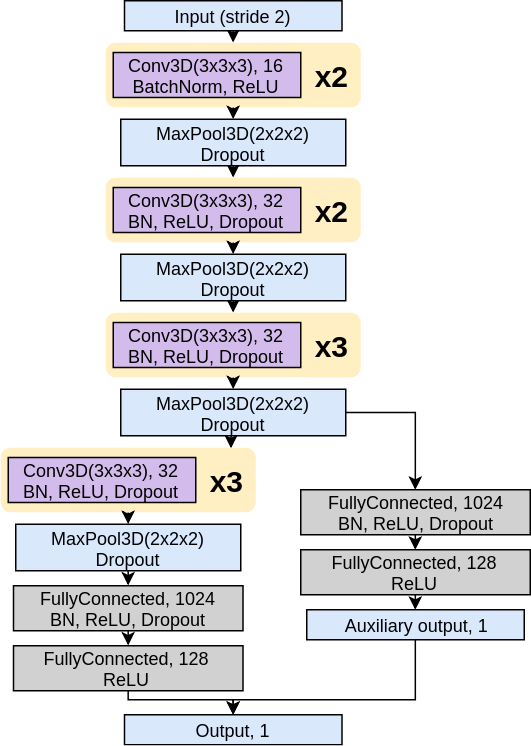}
        }
  \caption{VoxCNN model architectures used for fluid target prediction.}
  \label{fig:bratelli}
\end{figure}



\section{Experimental results}
\label{results}


In Table \ref{table:1} represented deep neural network architectures used and corresponding results for fluid intelligence prediction. Here the brain morphemic characteristics predictive capacity is considered as a baseline for prediction.

\begin{table}
\centering
\begin{tabular}{||l| c| c||}  \hline 
\# &  Model architecture & MSE \\ [2ex] 
\hline
 1 & Brain morphometry & 71.293\\ [1ex]
 2 & VoxCNN on brain T1 imagery & 71.777\\ [1ex]
 3 & VoxCNN on 3D segmented brain mask & 72.094\\ [1ex]
 4 & Ensemble: VoxCNNs on T1 and segmented mask & 71.314\\[1ex]
 5 & Ensemble: VoxCNNs on T1, segmented mask with morphology features & 70.635 \\ [2ex] \hline 
\end{tabular}

\caption{Model architectures and results on the  \texttt{Validation set}.}
\label{table:1}
\end{table}

\begin{table}
\centering
\begin{tabular}{||l| c| c||}  \hline 
\# &  Model architecture & MSE \\ [2ex] 
\hline
 1 & Ensemble: VoxCNNs on T1 and segmented mask & 92.8378\\[1ex]
 2 & Ensemble: VoxCNNs on T1, segmented mask with morphology features & 94.0808 \\ [2ex] \hline 
\end{tabular}

\caption{Model architectures and results for the fluid intelligence prediction on the \texttt{Test set}.}
\label{table:2}
\end{table}

The most accurate prediction (in terms of \texttt{MSE} on the validation set) was obtained as a weighted average of the two predictions by VoxCNN neural networks trained on different parts of the training sample:
\begin{enumerate}
\item VoxCNN network, trained on both brain T1 images and segmented images,
\item VoxCNN network (with auxiliary head for better convergence), trained on brain T1 images, segmented images and additional socio-demographic data. We used segmented brain masks and full brain imagery after pre-processing.
\end{enumerate} 

As a result, the first and the second network architectures showed $71.777$ and $71.094$ \texttt{MSE} scores on the \texttt{Validation} set. After averaging the predictions with adjusted weights $\frac{2}{3}$ and $\frac{1}{3}$, the final validation performance reached $70.635$ MSE when using ensembles of models. 

Then on the \texttt{Test} set the ensemble models yielded $92.8378$ and $94.0808$ \texttt{MSE} scores correspondingly.


\section{Discussion}
\label{discussion}

All constructed regression models provided \texttt{MSE}, which is equal approximately to $70$. These results are comparable to the baseline result, calculated using morphological characteristics on the \texttt{Validation} set. 

This incremental improvement and rather high errors across all models could potentially imply both the study design and the data inconsistency: the reason may be that structural T1-weighted images alone are not enough to predict fluid intelligence scores; at the same time brain functional data like fMRI might have more predictive power for cognitive assessment.

The top performing model was the combination  (a weighted average prediction) of two VoxCNN neural networks trained on different parts of the training sample, highlighting the potential strength of the  models' ensembles yielded $70.635$ \texttt{MSE} on the \texttt{Validation} set and $92.635$ \texttt{MSE} on the \texttt{Test} set.

\section{Conclusion}
\label{conclusion}

In our work for the first time ensembles of VoxCNN networks were applied to the 3D brain imagery regression task. According to the results of this architecture we could consider it as a consistent predictive tool for large datasets with heavy and multi-modal inputs. 

Due to the rich structure of the considered dataset there is enough room for further improvements.
A future work on the model hyperparameters optimization is needed in order to achieve better network convergence. We can use advanced approaches to initialization of neural network parameters \cite{ANNInit2016} and construction of ensembles \cite{Ensembles2013}. Sparse 3D convolutions could decrease memory requirements \cite{3DCNN2018}.

Transfer learning and domain adaptation techniques could potentially show better performance \cite{ghafoorian2017transfer,lu2017unsupervised,goetz2016dalsa}. Also we can utilize multi-fidelity approaches when solving the regression problem with multi-modal data \cite{MFGP2015,MFGP2017,MinimaxMFGP2017}.  Conformal prediction framework \cite{ConformalDR,VovkConformal2014,ConformalKRR2016} is a ready-to-use tool to assess prediction uncertainty.

The considered problem was formulated in the scope of the Project ``Machine Learning and Pattern Recognition for the development of diagnostic and clinical prognostic prediction tools in psychiatry, borderline mental disorders, and neurology'' (a part of the Skoltech Biomedical Initiative program).

\subsubsection{Acknowledgements}
The work was supported by the Russian Science Foundation under Grant 19-41-04109.


\bibliographystyle{alpha}

\bibliography{mybibliography}

\newcommand{\etalchar}[1]{$^{#1}$}
\begin{thebibliography}{HAGEB16}

\bibitem[ATS09]{avants2009advanced}
Brian~B Avants, Nick Tustison, and Gang Song.
\newblock Advanced normalization tools (ants).
\newblock {\em Insight j}, 2:1--35, 2009.

\bibitem[ATW{\etalchar{+}}11]{avants2011open}
Brian~B Avants, Nicholas~J Tustison, Jue Wu, Philip~A Cook, and James~C Gee.
\newblock An open source multivariate framework for n-tissue segmentation with
  evaluation on public data.
\newblock {\em Neuroinformatics}, 9(4):381--400, 2011.

\bibitem[BBT{\etalchar{+}}15]{brown2015national}
Sandra~A Brown, Ty~Brumback, Kristin Tomlinson, Kevin Cummins, Wesley~K
  Thompson, Bonnie~J Nagel, Michael~D De~Bellis, Stephen~R Hooper, Duncan~B
  Clark, Tammy Chung, et~al.
\newblock The national consortium on alcohol and neurodevelopment in
  adolescence (ncanda): a multisite study of adolescent development and
  substance use.
\newblock {\em Journal of studies on alcohol and drugs}, 76(6):895--908, 2015.

\bibitem[BE16]{ANNInit2016}
E.~Burnaev and P.~Erofeev.
\newblock The influence of parameter initialization on the training time and
  accuracy of a nonlinear regression model.
\newblock {\em Journal of Communications Technology and Electronics},
  61(6):646--660, Jun 2016.

\bibitem[BN16]{ConformalKRR2016}
E.~Burnaev and I.~Nazarov.
\newblock Conformalized kernel ridge regression.
\newblock In {\em 2016 15th IEEE International Conference on Machine Learning
  and Applications (ICMLA)}, pages 45--52, 2016.

\bibitem[BP13]{Ensembles2013}
E.~V. Burnaev and P.~V. Prikhod'ko.
\newblock On a method for constructing ensembles of regression models.
\newblock {\em Automation and Remote Control}, 74(10):1630--1644, Oct 2013.

\bibitem[BV14]{VovkConformal2014}
E.~Burnaev and V.~Vovk.
\newblock Efficiency of conformalized ridge regression.
\newblock In Maria~Florina Balcan, Vitaly Feldman, and Csaba Szepesvari,
  editors, {\em Proceedings of The 27th Conference on Learning Theory},
  volume~35 of {\em Proceedings of Machine Learning Research}, pages 605--622,
  Barcelona, Spain, 13--15 Jun 2014. PMLR.

\bibitem[BZ15]{MFGP2015}
E.~Burnaev and A.~Zaytsev.
\newblock Surrogate modeling of multifidelity data for large samples.
\newblock {\em Journal of Communications Technology and Electronics},
  60(12):1348--1355, 2015.

\bibitem[Car93]{carroll_1993}
John~B. Carroll.
\newblock {\em Human Cognitive Abilities: A Survey of Factor-Analytic Studies}.
\newblock Cambridge University Press, 1993.

\bibitem[CDY{\etalchar{+}}18]{chen2018voxresnet}
Hao Chen, Qi~Dou, Lequan Yu, Jing Qin, and Pheng-Ann Heng.
\newblock Voxresnet: Deep voxelwise residual networks for brain segmentation
  from 3d mr images.
\newblock {\em NeuroImage}, 170:446--455, 2018.

\bibitem[Cox96]{cox1996afni}
Robert~W Cox.
\newblock Afni: software for analysis and visualization of functional magnetic
  resonance neuroimages.
\newblock {\em Computers and Biomedical research}, 29(3):162--173, 1996.

\bibitem[CPC16]{canziani2016analysis}
Alfredo Canziani, Adam Paszke, and Eugenio Culurciello.
\newblock An analysis of deep neural network models for practical applications.
\newblock {\em arXiv preprint arXiv:1605.07678}, 2016.

\bibitem[ESH19]{eckle2019comparison}
Konstantin Eckle and Johannes Schmidt-Hieber.
\newblock A comparison of deep networks with relu activation function and
  linear spline-type methods.
\newblock {\em Neural Networks}, 110:232--242, 2019.

\bibitem[GMK{\etalchar{+}}17]{ghafoorian2017transfer}
Mohsen Ghafoorian, Alireza Mehrtash, Tina Kapur, Nico Karssemeijer, Elena
  Marchiori, Mehran Pesteie, Charles~RG Guttmann, Frank-Erik de~Leeuw, Clare~M
  Tempany, Bram van Ginneken, et~al.
\newblock Transfer learning for domain adaptation in mri: Application in brain
  lesion segmentation.
\newblock In {\em International Conference on Medical Image Computing and
  Computer-Assisted Intervention}, pages 516--524. Springer, 2017.

\bibitem[GWB{\etalchar{+}}16]{goetz2016dalsa}
Michael Goetz, Christian Weber, Franciszek Binczyk, Joanna Polanska, Rafal
  Tarnawski, Barbara Bobek-Billewicz, Ullrich Koethe, Jens Kleesiek, Bram
  Stieltjes, and Klaus~H Maier-Hein.
\newblock Dalsa: domain adaptation for supervised learning from sparsely
  annotated mr images.
\newblock {\em IEEE transactions on medical imaging}, 35(1):184--196, 2016.

\bibitem[HAGEB16]{hosseini2016alzheimer}
Ehsan Hosseini-Asl, Georgy Gimel'farb, and Ayman El-Baz.
\newblock Alzheimer's disease diagnostics by a deeply supervised adaptable 3d
  convolutional network.
\newblock {\em arXiv preprint arXiv:1607.00556}, 2016.

\bibitem[HHM{\etalchar{+}}18]{hagler2018image}
Donald~J Hagler, Sean~N Hatton, Carolina Makowski, M~Daniela Cornejo, Damien~A
  Fair, Anthony~Steven Dick, Matthew~T Sutherland, BJ~Casey, Deanna~M Barch,
  Michael~P Harms, et~al.
\newblock Image processing and analysis methods for the adolescent brain
  cognitive development study.
\newblock {\em bioRxiv}, page 457739, 2018.

\bibitem[HZRS16]{he2016deep}
Kaiming He, Xiangyu Zhang, Shaoqing Ren, and Jian Sun.
\newblock Deep residual learning for image recognition.
\newblock In {\em Proceedings of the IEEE conference on computer vision and
  pattern recognition}, pages 770--778, 2016.

\bibitem[ILTT11]{iglesias2011robust}
Juan~Eugenio Iglesias, Cheng-Yi Liu, Paul~M Thompson, and Zhuowen Tu.
\newblock Robust brain extraction across datasets and comparison with publicly
  available methods.
\newblock {\em IEEE transactions on medical imaging}, 30(9):1617--1634, 2011.

\bibitem[ISA{\etalchar{+}}18]{DepressionAWE2018}
S.~Ivanov, M.~Sharaev, A.~Artemov, E.~Kondratyeva, A.~Cichocki,
  S.~Sushchinskaya, E.~Burnaev, and A.~Bernstein.
\newblock Learning connectivity patterns via graph kernels for fmri-based
  depression diagnostics.
\newblock In {\em Proc. of IEEE International Conference on Data Mining
  Workshops (ICDMW)}, pages 308--314, 2018.

\bibitem[KBB18]{ConformalDR}
A.~Kuleshov, A.~Bernstein, and E.~Burnaev.
\newblock Conformal prediction in manifold learning.
\newblock In Alex Gammerman, Vladimir Vovk, Zhiyuan Luo, Evgueni Smirnov, and
  Ralf Peeters, editors, {\em Proceedings of the Seventh Workshop on Conformal
  and Probabilistic Prediction and Applications}, volume~91 of {\em Proceedings
  of Machine Learning Research}, pages 234--253. PMLR, 11--13 Jun 2018.

\bibitem[KSBD17]{korolev2017residual}
Sergey Korolev, Amir Safiullin, Mikhail Belyaev, and Yulia Dodonova.
\newblock Residual and plain convolutional neural networks for 3d brain mri
  classification.
\newblock In {\em 2017 IEEE 14th International Symposium on Biomedical Imaging
  (ISBI 2017)}, pages 835--838. IEEE, 2017.

\bibitem[LZC{\etalchar{+}}17]{lu2017unsupervised}
Hao Lu, Lei Zhang, Zhiguo Cao, Wei Wei, Ke~Xian, Chunhua Shen, and Anton
  van~den Hengel.
\newblock When unsupervised domain adaptation meets tensor representations.
\newblock In {\em Proceedings of the IEEE International Conference on Computer
  Vision}, pages 599--608, 2017.

\bibitem[MNA16]{milletari2016v}
Fausto Milletari, Nassir Navab, and Seyed-Ahmad Ahmadi.
\newblock V-net: Fully convolutional neural networks for volumetric medical
  image segmentation.
\newblock In {\em 2016 Fourth International Conference on 3D Vision (3DV)},
  pages 565--571. IEEE, 2016.

\bibitem[NKB18]{3DCNN2018}
A.~Notchenko, Ye. Kapushev, and E.~Burnaev.
\newblock Large-scale shape retrieval with sparse 3d convolutional neural
  networks.
\newblock In Wil~M.P. van~der Aalst, D.~Ignatov, M.~Khachay, and et~al.,
  editors, {\em Analysis of Images, Social Networks and Texts}, pages 245--254,
  Cham, 2018. Springer International Publishing.

\bibitem[PAS{\etalchar{+}}18]{pominova2018voxelwise}
Marina Pominova, Alexey Artemov, Maksim Sharaev, Ekaterina Kondrateva,
  Alexander Bernstein, and Evgeny Burnaev.
\newblock Voxelwise 3d convolutional and recurrent neural networks for epilepsy
  and depression diagnostics from structural and functional mri data.
\newblock In {\em 2018 IEEE International Conference on Data Mining Workshops
  (ICDMW)}, pages 299--307. IEEE, 2018.

\bibitem[PKB{\etalchar{+}}17]{pfefferbaum2017altered}
Adolf Pfefferbaum, Dongjin Kwon, Ty~Brumback, Wesley~K Thompson, Kevin Cummins,
  Susan~F Tapert, Sandra~A Brown, Ian~M Colrain, Fiona~C Baker, Devin Prouty,
  et~al.
\newblock Altered brain developmental trajectories in adolescents after
  initiating drinking.
\newblock {\em American journal of psychiatry}, 175(4):370--380, 2017.

\bibitem[PLN{\etalchar{+}}16]{paul2016dissociable}
Erick~J Paul, Ryan~J Larsen, Aki Nikolaidis, Nathan Ward, Charles~H Hillman,
  Neal~J Cohen, Arthur~F Kramer, and Aron~K Barbey.
\newblock Dissociable brain biomarkers of fluid intelligence.
\newblock {\em NeuroImage}, 137:201--211, 2016.

\bibitem[RZSP10]{rohlfing2010sri24}
Torsten Rohlfing, Natalie~M Zahr, Edith~V Sullivan, and Adolf Pfefferbaum.
\newblock The sri24 multichannel atlas of normal adult human brain structure.
\newblock {\em Human brain mapping}, 31(5):798--819, 2010.

\bibitem[SAA{\etalchar{+}}18]{Pipeline2018}
M.~Sharaev, A.~Andreev, A.~Artemov, E.~Burnaev, E.~Kondratyeva,
  S.~Sushchinskaya, I.~Samotaeva, V.~Gaskin, and A.~Bernstein.
\newblock Pattern recognition pipeline for neuroimaging data.
\newblock In Luca Pancioni, Friedhelm Schwenker, and Edmondo Trentin, editors,
  {\em Artificial Neural Networks in Pattern Recognition}, pages 306--319,
  Cham, 2018. Springer International Publishing.

\bibitem[SAK{\etalchar{+}}18]{Epilepsy2018}
M.~Sharaev, A.~Artemov, E.~Kondratyeva, S.~Sushchinskaya, E.~Burnaev,
  A.~Bernstein, R.~Akzhigitov, and A.~Andreev.
\newblock Mri-based diagnostics of depression concomitant with epilepsy: in
  search of the potential biomarkers.
\newblock In {\em Proceedings of IEEE 5th International Conference on Data
  Science and Advanced Analytics}, pages 555--564, 2018.

\bibitem[SLJ{\etalchar{+}}15]{szegedy2015going}
Christian Szegedy, Wei Liu, Yangqing Jia, Pierre Sermanet, Scott Reed, Dragomir
  Anguelov, Dumitru Erhan, Vincent Vanhoucke, and Andrew Rabinovich.
\newblock Going deeper with convolutions.
\newblock In {\em Proceedings of the IEEE conference on computer vision and
  pattern recognition}, pages 1--9, 2015.

\bibitem[Smi02]{smith2002fast}
Stephen~M Smith.
\newblock Fast robust automated brain extraction.
\newblock {\em Human brain mapping}, 17(3):143--155, 2002.

\bibitem[SVI{\etalchar{+}}16]{szegedy2016rethinking}
Christian Szegedy, Vincent Vanhoucke, Sergey Ioffe, Jon Shlens, and Zbigniew
  Wojna.
\newblock Rethinking the inception architecture for computer vision.
\newblock In {\em Proceedings of the IEEE conference on computer vision and
  pattern recognition}, pages 2818--2826, 2016.

\bibitem[SZ14]{simonyan2014very}
Karen Simonyan and Andrew Zisserman.
\newblock Very deep convolutional networks for large-scale image recognition.
\newblock {\em arXiv preprint arXiv:1409.1556}, 2014.

\bibitem[SZCZ10]{sadananthan2010skull}
Suresh~A Sadananthan, Weili Zheng, Michael~WL Chee, and Vitali Zagorodnov.
\newblock Skull stripping using graph cuts.
\newblock {\em NeuroImage}, 49(1):225--239, 2010.

\bibitem[TGJ{\etalchar{+}}15]{tompson2015efficient}
Jonathan Tompson, Ross Goroshin, Arjun Jain, Yann LeCun, and Christoph Bregler.
\newblock Efficient object localization using convolutional networks.
\newblock In {\em Proceedings of the IEEE Conference on Computer Vision and
  Pattern Recognition}, pages 648--656, 2015.

\bibitem[ZB17a]{MFGP2017}
A.~Zaytsev and E.~Burnaev.
\newblock Large scale variable fidelity surrogate modeling.
\newblock {\em Annals of Mathematics and Artificial Intelligence},
  81(1):167--186, Oct 2017.

\bibitem[ZB17b]{MinimaxMFGP2017}
A.~Zaytsev and E.~Burnaev.
\newblock {Minimax Approach to Variable Fidelity Data Interpolation}.
\newblock In Aarti Singh and Jerry Zhu, editors, {\em Proceedings of the 20th
  International Conference on Artificial Intelligence and Statistics},
  volume~54 of {\em Proceedings of Machine Learning Research}, pages 652--661,
  Fort Lauderdale, FL, USA, 20--22 Apr 2017. PMLR.

\bibitem[ZLL18]{zhu2018prediction}
Meifang Zhu, Bing Liu, and Jin Li.
\newblock Prediction of general fluid intelligence using cortical measurements
  and underlying genetic mechanisms.
\newblock In {\em IOP Conference Series: Materials Science and Engineering},
  volume 381, page 012186. IOP Publishing, 2018.

\end{thebibliography}

\end{document}